\documentclass[aps,prl,twocolumn,showpacs,floatfix]{revtex4-1}

\usepackage{graphicx}% Include figure files
\usepackage{amssymb}
\usepackage{natbib}
\usepackage{color}
\begin{document}

%\preprint{}

\title{Chain Retraction in Highly Entangled Stretched Polymer Melts}

\author{Hsiao-Ping Hsu}\email{hsu@mpip-mainz.mpg.de}
\author{Kurt Kremer}\email{kremer@mpip-mainz.mpg.de}
\affiliation{Max-Planck-Institut f\"ur Polymerforschung, Ackermannweg 10, 55128, Mainz, Germany}
%\affiliation{Max-Planck-Institut f\"ur Polymerforschung, Ackermannweg 10, 55128, Mainz}

%\date{\today}

\begin{abstract}
We use computer simulations to
study the relaxation of strongly deformed highly entangled polymer melts in
the nonlinear viscoelastic regime, focusing on anisotropic chain conformations 
after isochoric elongation.
The Doi-Edwards tube model and its Graham-Likhtman-McLeish-Milner (GLaMM) 
extension, incorporating contour
length fluctuation and convective constraint release, predict a retraction 
of the polymer chain extension in all directions, setting in immediately
after deformation. This prediction has been challenged by experiment, 
simulation, and other theoretical studies, questioning the general validity of
the tube concept. 
For very long chains we observe the initial contraction of the chain extension
parallel and perpendicular to the stretching direction. 
However, the effect is significantly weaker than predicted by the GLaMM model.  
We also show that the first anisotropic term of an expansion of 
the 2D scattering function qualitatively agrees to predictions of 
the GLaMM model, providing an option for direct experimental tests.
\end{abstract}

\pacs{83.80Sg, 83.50.-v, 83.10.Rs}

%83.80.Sg 	Polymer melts
%83.50.-v 	Deformation and flow
%83.10.Rs 	Computer simulation of molecular and particle dynamics

%\keywords{}

\maketitle

The reptation model and its extensions, based on conformational properties and 
entanglement effects in dense polymer systems, represent the basis of our 
current understanding of viscoelastic properties of modern polymer materials, 
which are omnipresent in our daily life products and in technology. 
In the linear viscoelastic regime, the theory originally developed by Doi, 
Edwards, and de 
Gennes~\cite{deGennes1979,Doi1980,Doi1981,Doi1983,Doi1986,Rubinstein2003} based
on the original tube concept of Edwards~\cite{Edwards1967} successfully 
describes dynamics and viscoelasticity, e.g., stress relaxation, of highly 
entangled polymer melts. It is strongly supported in detail by 
simulation~\cite{Kremer1988,Kremer1990,Paul1991,Wittmer1992,Kremer1992,
Kopf1997,Puetz2000,Harmandaris2003,Tsolou2005,Hsu2016,Hsu2017} 
and experiment~\cite{Fetters1994,Richter2006,Graessley2008,Herrmann2012}. 
To account for finite chain length corrections, refinements of the original 
concept have been developed, namely  the effect of contour length fluctuation
(CLF)~\cite{Doi1983,Lodge1999,Likhtman2002,Abdel2004,Likhtman2014,Furtado2014}, 
and constraint release (CR)~\cite{Likhtman2002, Klein1978, Daoud1979, 
Rubinstein1988,Milner2001,McLeish2002,McLeish2003}. 
These modify pure reptation and correctly reproduce the disentanglement time of
$\tau_{d,N} \propto N^{3.4}$, $N$ being the degree of polymerization of the 
chains. In the nonlinear viscoelastic regime, Doi and Edwards~\cite{Doi1986} 
assume that polymer chains in a melt deform affinely along the chain contour, 
following the sample deformation. 
On scales above the tube diameter this has recently been confirmed by us~\cite{Hsu2018a}.
The chain radius of gyration along the stretching direction increases and 
simultaneously decreases in the perpendicular direction. 
Immediately after deformation - still within the affinely deformed tube - the 
stress along the contour of the chain causes an initial retraction along the 
tube. \textit{All} linear dimensions of deformed chains are expected to first 
decrease, while the chains try to retract back into the tube.
Following the refined Graham-Likhtman-McLeish-Milner
(GLaMM)  tube model~\cite{Graham2003} [includes CLF, CR, and
convective constraint release (CCR)] this initial retraction is expected to 
last for up to the Rouse time of the chains. 
However, it is not clear in which way these concepts apply to local 
conformational properties and the nonlinear viscoelasticity of polymer melts. 

Based on neutron scattering experiments of highly stretched polystyrene melts 
Wang {\it et al.}~\cite{Wang2017} question the validity of the whole tube concept. 
By careful analysis of two-dimensional anisotropic small-angle neutron 
scattering (SANS) spectra of polymer melts having $Z=N/N_e=34$ entanglements 
per chain ($N_e$ being the entanglement length),
they could not observe the predicted initial chain retraction.
Their subsequent molecular dynamics simulation~\cite{Xu2018} of a standard, 
fully flexible bead spring model of polymer melts~\cite{Kremer1990,Kremer1992}
of $Z=33$ supports their experimental findings
(taking $N_e \approx 85$, as estimated through a primitive path 
analysis~\cite{Hoy2009,Moreira2015},  $Z \approx 24$).
In contrast, earlier work on nonlinear 
rheology of highly entangled polymer 
melts~\cite{Bent2003,Blanchard2005,Graham2006} 
supports the theoretical prediction of chain retraction by SANS.
\textcolor{black}{In Refs~\cite{Muller1992,Muller1988,Bent2003,Graham2006} the authors 
observe clear signatures of anisotropically deformed conformations of monodisperse entangled 
polystyrene melts in nonlinear flow, and even for unentangled chains subject to extremely fast flow~\cite{Kroeger1997}. In Refs.~\cite{Bent2003,Graham2006} the authors
also show the subsequent relaxation in 
agreement with the GLaMM tube model}. In Ref.~\cite{Blanchard2005}, 
Blanchard {\it et al.} observe a minimum in the deformed radius of gyration 
perpendicular to the stretching direction after cessation of flow for long, 
well-entangled polyisoprene chains of $Z=58$. In view of these contradictory 
results we present a study of the conformational relaxation behavior of polymer 
melts right after a large step elongation for different numbers of 
entanglements per chain ranging from about $Z=18$ to $Z=72$. 
By comparing chain conformations and an expansion of small angle scattering 
patterns~\cite{Wang2017,Xu2018} in \textcolor{black}{spherical harmonics~\cite{Lindner1989}} 
we demonstrate how the 
overall scattering patterns infer the internal structure of the chain 
conformations.

\begin{figure*}[t!]
\begin{center}
(a)\includegraphics[width=0.32\textwidth,angle=270]{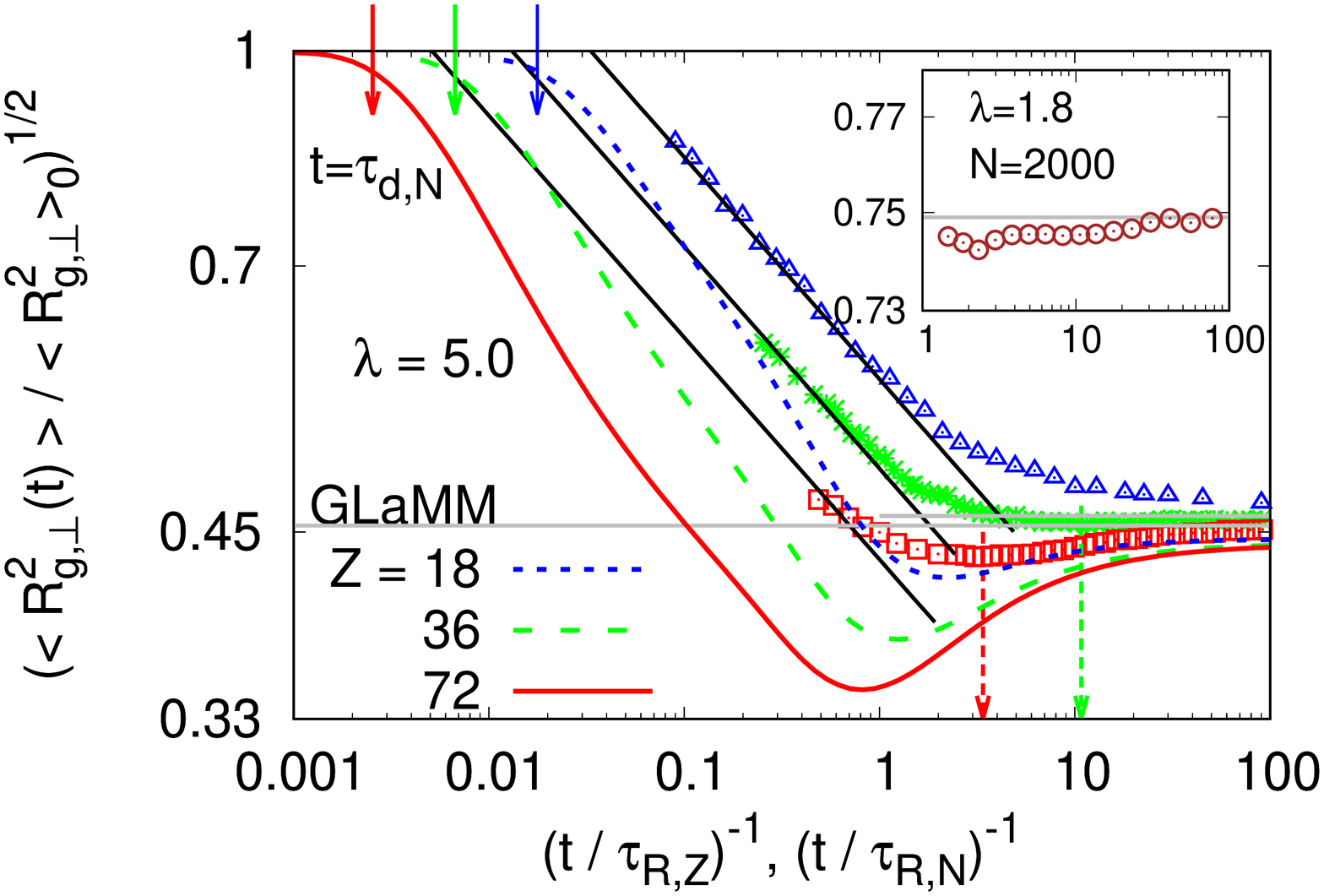} 
\hspace{0.1cm}
(b)\includegraphics[width=0.32\textwidth,angle=270]{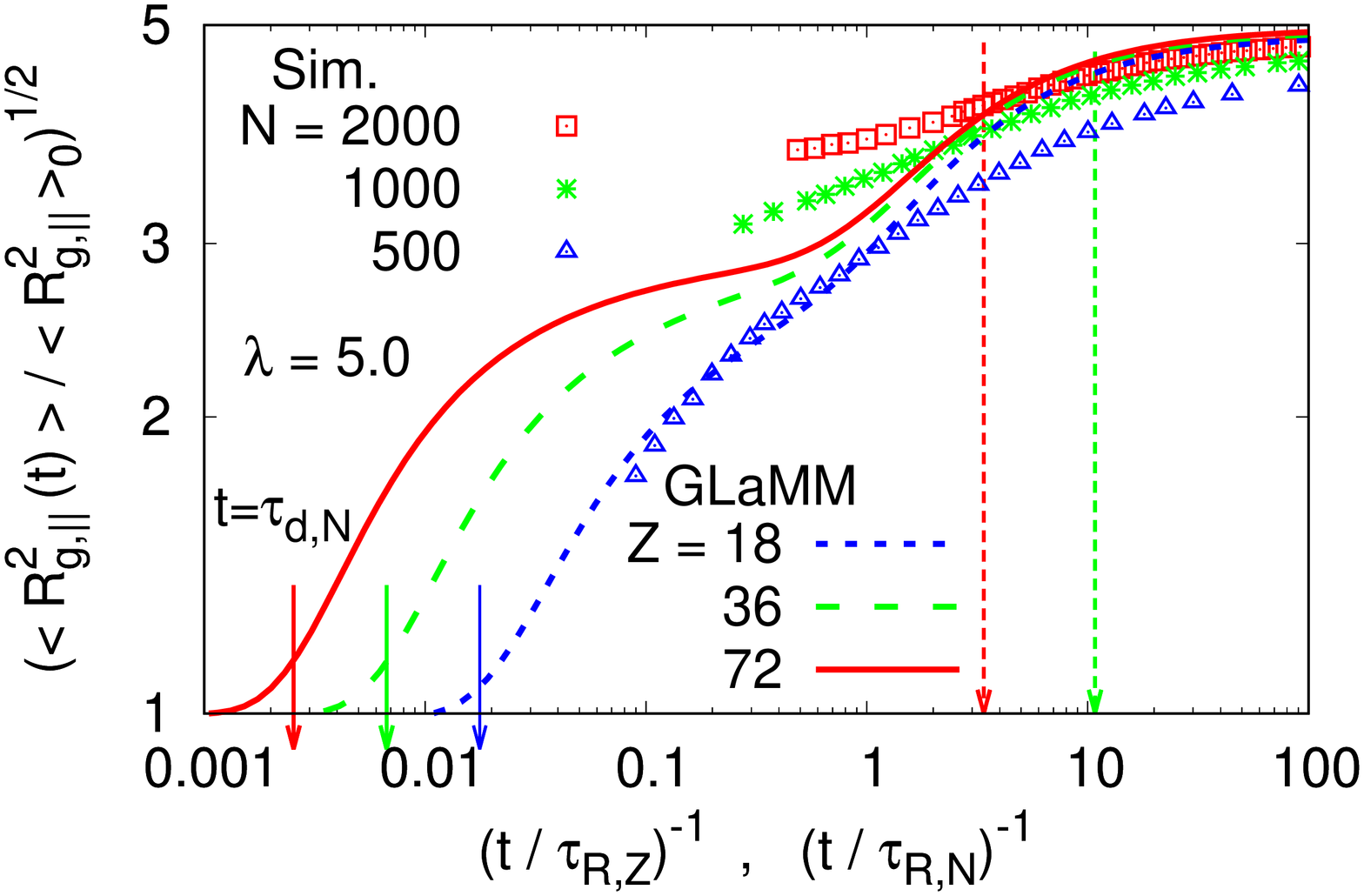}\\
\caption{Log-log plot of the rescaled root mean square radius of gyration
in the direction perpendicular and parallel to the stretching direction,
$(\langle R_{g,\perp}^2 (t)\rangle/\langle R_{g,\perp}^2 \rangle_0)^{1/2}$ (a)
and 
$(\langle R_{g,||}^2 (t)\rangle/\langle R_{g,||}^2 \rangle_0)^{1/2}$ (b),
respectively, plotted versus inverse rescaled time,
$(t/\tau_{R,N})^{-1}$, for 
$N=500$, $1000$, and $2000$, as indicated. The corresponding disentanglement 
times $t=\tau_{d,N}$ are pointed by solid arrows near one on the $y$ axis.
Theoretical predictions from the GLaMM model~\cite{Graham2003} versus 
$(t/\tau_{R,Z})^{-1}$ are shown for comparison. 
Minimum values of
$\langle R_{g,\perp}^2 (t)\rangle/\langle R_{g,\perp}^2 \rangle_0$
are marked by dashed arrows,
which also indicate the onset of relaxation delay in (b).
The horizontal line in (a) gives  
$(\langle R_{g,\perp}^2 (t)\rangle/\langle R_{g,\perp}^2 \rangle_0)^{1/2}$
right after elongation for $N=1000$ and $2000$.
Straight lines ($a_gx^{-b_g}$) indicate best fits to our simulation data
for $(\tau_{R,N}/t)<1.0$ (cf. text).
For comparison data for $N=2000$ at $\lambda=1.8$ are shown in the inset 
of (a).}
\label{fig-rel-Rgxyz}
\end{center}
\end{figure*}

We have performed extensive molecular dynamics (MD) simulations of 
strongly deformed polymer melts~\cite{Hsu2018a} using the ESPResSo++ 
package~\cite{Espressopp}. \textcolor{black}{Starting from fully equilibrated melts of highly 
entangled bead-spring chains with a weak bond bending constant of 
$k_\theta=1.5\epsilon$~\cite{Kremer1990, Zhang2014, Moreira2015, Hsu2016, 
Hsu2017} chains behave as ideal chains above the Kuhn length corresponding to
  $ 2.8(1) \ell_b$, $\ell_b \approx 0.964\sigma$ being the bond length. 
Isothermal MD simulations at temperature $T=1\epsilon/k_B$ have been performed 
(for details see the Supplemental Material
in Ref.~\cite{Hsu2018a}). Lennard-Jones energy, time, and length units 
 denoted by $\epsilon$, $\tau$, and $\sigma$, respectively, are used throughout this work.}
For these parameters the entanglement length corresponds to $N_e = 28$ 
monomers~\cite{Moreira2015,Hsu2016}.
We apply an isochoric and uniaxial 
elongation along the $x$ direction with a deformation rate 
$\dot{\varepsilon}=77\tau_{R,N}^{-1}=(77/Z^2)\tau_{e}^{-1}$,
i.e. $\tau_{R,N}^{-1} < \dot{\varepsilon} < \tau_{e}^{-1}$,
up to a total strain of  $\lambda=L_x/L_0=5$.
Here $\tau_{R,N}=\tau_0N^2$ with $\tau_0=2.89\tau$ 
and ${\tau_e}=\tau_0N_e^2$ are Rouse 
times of a chain of length $N$ and of an entanglement length $N_e$, 
respectively.
This is the relevant nonlinear viscoelastic regime, where a delicate 
interplay between deformation rate and internal conformational relaxation 
plays a crucial role~\cite{Likhtman2002,McLeish2002,McLeish2003,Graham2003}.
For times up to about the Rouse time of the chains we have seen that the 
relaxation along the tube is by no means homogeneous; i.e., the primitive 
paths~\cite{Everaers2004,Sukumaran2005,Everaers2012} exhibit long-lived 
clustering of topological constraints in the deformed state, leading to 
significantly delayed relaxation~\cite{Hsu2018a}.
This is not accounted for by any of the current theoretical concepts. 
Here, however, we focus on the initial relaxation of experimentally more 
directly accessible global conformational properties of deformed polymer melts,
where observed deviations from the GLaMM model~\cite{Graham2003} have been 
taken to question the validity of the tube concept as a 
whole~\cite{Wang2017,Xu2018,Blanchard2005}.

Subject to uniaxial elongation the average conformation of single chains in 
a melt exhibits axial symmetry along the stretching direction ($x$ axis).
Therefore, the mean square radius of gyration which describes the chain 
conformations should be decomposed into two components parallel and 
perpendicular to the stretching direction, i.e.
$\langle R_g^2 \rangle = \langle R_{g,||}^2 \rangle + 
\langle R_{g,\perp}^2 \rangle$, 
\textcolor{black}{and $\langle R_{g,||}^2 \rangle=\langle R_{g,\perp}^2 \rangle/2$ in equilibrium.}

Based on the tube model one would expect an overdamped initial retraction 
process in \textit{both} directions parallel \textit{and} perpendicular to the 
$x$ axis~\cite{Doi1986,Graham2003}. While this is obvious for the extension 
parallel to the stretching direction, this effect is expected to be much 
weaker for the perpendicular one, as it eventually has to turn and increase 
towards the equilibrium value.
Time evolution of the rescaled two components of radius of gyration, 
$(\langle R_{g,\perp}^2 \rangle/\langle R_{g,\perp}^2 \rangle_0)^{1/2}$  
and $(\langle R_{g,||}^2 \rangle/\langle R_{g,||}^2 \rangle_0)^{1/2}$, 
for single chains of sizes $N=500$, $1000$, and $2000$ ($Z \approx18$, $36$, 
and $72$) in melts during relaxation are shown in Fig.~\ref{fig-rel-Rgxyz} and 
compared to the GLaMM model~\cite{Graham2003,Graham2013} (see Supplemental Material~\cite{SM}).
The symbols $\langle \cdots \rangle$ and $\langle \cdots \rangle_0$ stand for 
the average over $n_c=1000$ chains in deformed and unperturbed (i.e., fully 
equilibrated) polymer melts, respectively.
The parameters $c_\nu=0.1$ and $R_s=2.0$
are set to the same values as they 
were tested in the GLaMM model~\cite{Graham2003}.
Except for 
$(\langle R_{g,\perp}^2 \rangle/\langle R_{g,\perp}^2\rangle_0)^{1/2}$ of the 
shortest chains of 
$N=500$ (i.e., $Z \approx 18$), we see that \textit{both} components of $R_g$ 
for deformed polymer melts initially decrease with increasing relaxation time.
Evidently, our results qualitatively capture the signature of the initial chain
retraction mechanism~\cite{Doi1986,Graham2003} right after a large step 
elongation. 

\begin{figure*}[t!]
\begin{center}
(a)\includegraphics[width=0.32\textwidth,angle=270]{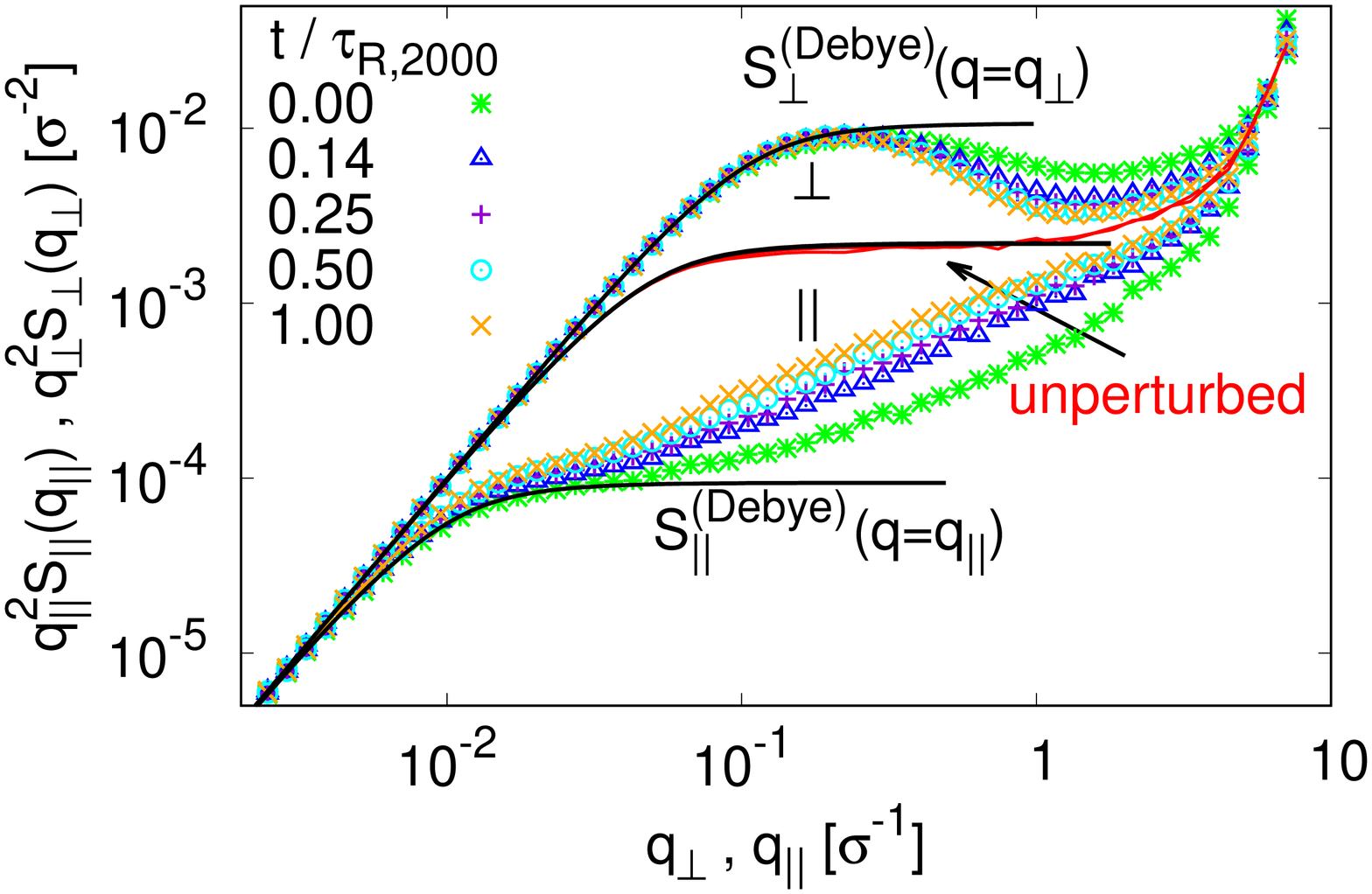}
\hspace{0.1cm}
(b)\includegraphics[width=0.32\textwidth,angle=270]{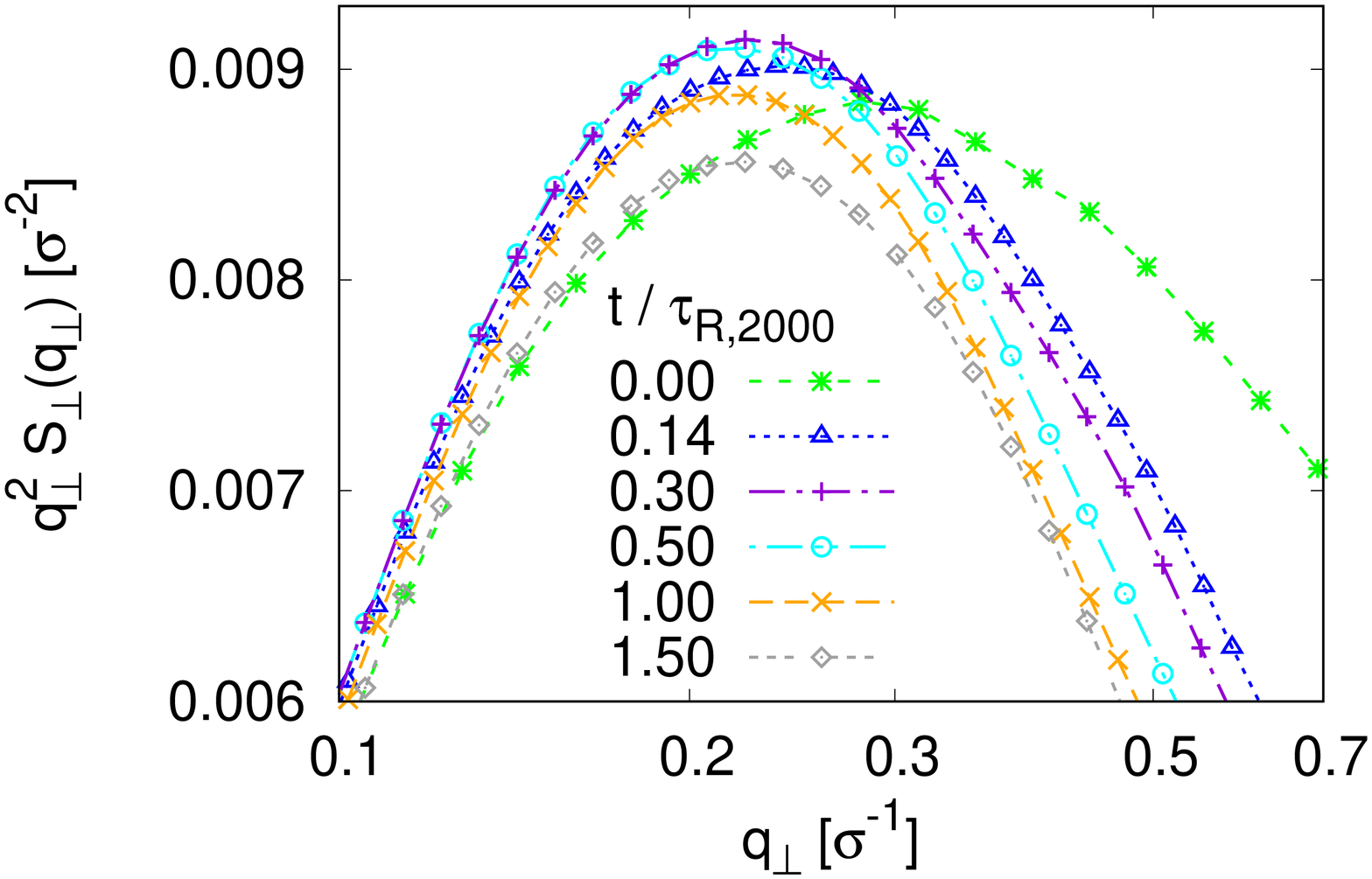}
\caption{Kratky-plot of the two components of the single
chain structure factor parallel and perpendicular to the stretching direction,
$q_{||}^2\textcolor{black}{S_{||}}(q_{||})$ and
$q_\perp^2\textcolor{black}{S_{\perp}}(q_\perp)$,
respectively (a), and $q_\perp^2\textcolor{black}{S_{\perp}}(q_\perp)$
for $0.1\sigma^{-1}<q_\perp<0.7\sigma^{-1}$ (b).
Data are for chains of size $N = 2000$.
Several values of the relaxation time $t/\tau_{R,N}$ are shown,
as indicated.
Data for the unperturbed polymer melt (red curve) and the decomposed
Debye functions~\cite{Benoit1975,Hsu2012}:
\textcolor{black}{$S_{\alpha}^{\rm (Debye)}(q=q_{\alpha}) = 2 \left [
\exp(-X_{\alpha})-1+X_{\alpha} \right ]/{X_{\alpha}^2}$ with
$X_{\alpha=||}=3q_{||}^2\langle R_{g,||}^2 \rangle$
and $X_{\alpha=\perp}=3q_{\perp}^2 \langle R_{\perp}^2 \rangle/2$
are also shown in (a) for comparison. Note that
$S_{||}^{\rm (Debye)}(q)=S_{\perp}^{\rm (Debye)}(q)$ for unperturbed
polymer melts.}}
\label{fig-rel-Sqcxyz}
\end{center}
\end{figure*}

Since the chains finally must relax to their equilibrium conformation,  
$(\langle R_{g,\perp}^2 \rangle/\langle R_{g,\perp}^2\rangle_0)^{1/2}$ 
will go through a minimum until it increases towards unity.
In Fig.~\ref{fig-rel-Rgxyz}(a) we see that 
$(\langle R_{g,\perp}^2 \rangle/\langle R_{g,\perp}^2\rangle_0)^{1/2}$ first 
decreases, reaches a minimum at $t/\tau_{R,1000}=0.09$ for $N=1000$ and 
$t/\tau_{R,2000}=0.30$ for $N=2000$, and then turns around and gradually 
increases. With increasing $Z$ the minimum becomes more pronounced and is 
shifted to later times, however still below $\tau_{R,N}$. 
The GLaMM model predicts a minimum at $t \approx \tau_{R,Z}$ and a \textcolor{black}{significantly stronger 
signature of retraction for the same values of $Z$, thus only qualitatively agreeing to our data.}
[Note that our data for $N=2000$ and $\lambda=1.8$ 
shown in Fig.~\ref{fig-rel-Rgxyz}(a) (inset) indicate the signature becomes 
much weaker with decreasing $\lambda$. From that it is not surprising that 
the minimum in $\langle R_{g,\perp}^2 \rangle$ has not been observed in 
Ref~\cite{Xu2018}].
\textcolor{black}{It is tempting to extrapolate the data to 
$(\langle R^2_{g,\perp} \rangle/\langle R^2_{g,\perp} \rangle_0)^{1/2}=1$ by  a fitting function $f(x)=a_gx^{-b_g}$ for $t>\tau_{R,N}$. For $N=500$ we estimate $b_g=0.16$ and $a_g=0.58$. Assuming the same power law for $N=1000$, $2000$, because all systems are deep in the entangled regime, we arrive at $a_g = 0.50$ and $0.43$, for $N=1000$, $2000$. By that we obtain equilibration time estimates of $t_{{\rm eq},N}[=(1/a_g)^{1/b_g}\tau_{R,N}]= 30\tau_{R,500}$,  $76\tau_{R,1000}$, and $195\tau_{R,2000}$, close to $\tau_{d,N}/2$, $\tau_{d,N}=(N/N_e)^{1.4}\tau_{R,N}$.}
For the GLaMM model, one obtains $t_{\textcolor{black}{\rm eq},Z}=50\tau_{R,Z=18}$,
$148\tau_{R,Z=36}$, and $363\tau_{R,Z=72}$ with the parameters 
$b_g=0.21$ and $a_g=0.44$, $0.35$, and $0.29$ for $Z=18$, $36$, and
$72$, respectively, based on $\tau_{d,Z}=Z^{1.4}\tau_{R,Z}$. 
The above assumes an unperturbed relaxation until isotropic 
chain conformations are reached. 
Though intuitive, this most probably cannot be the case, 
as indicated by the data for $\langle R_{g,||}^2 \rangle$, as well as by
previous primitive path analysis~\cite{Hsu2018a}. 
$(\langle R_{g,||}^2 \rangle/\langle R_{g,||}^2\rangle_0)^{1/2}$ decreases 
monotonically with time $t$ while the relaxation \textcolor{black}{rate still becomes smaller with time ($N = 1000$, $2000$).} 
Eventually we observe the signature of an intermediate plateau well 
above and significantly earlier than the regime predicted by GLaMM, pointing towards a 
significantly delayed conformational relaxation. 
This relaxation retardation of the deformed chains has been attributed to an 
inhomogeneous distribution of entanglement points along the primitive 
paths~\cite{Hsu2018a}, not accounted for in current theoretical 
models. A similar delay has been observed in the context of rheological 
experiments of very long, highly entangled polymer chains by several 
authors~\cite{Archer1995,Archer2002,Venerus2006}.
\textcolor{black}{The GLaMM model predicts the equilibrium melt disentanglement time of the chain to be the longest relaxation time. }

\begin{figure}[htb!]
\begin{center}
(a)\includegraphics[width=0.25\textwidth,angle=270]{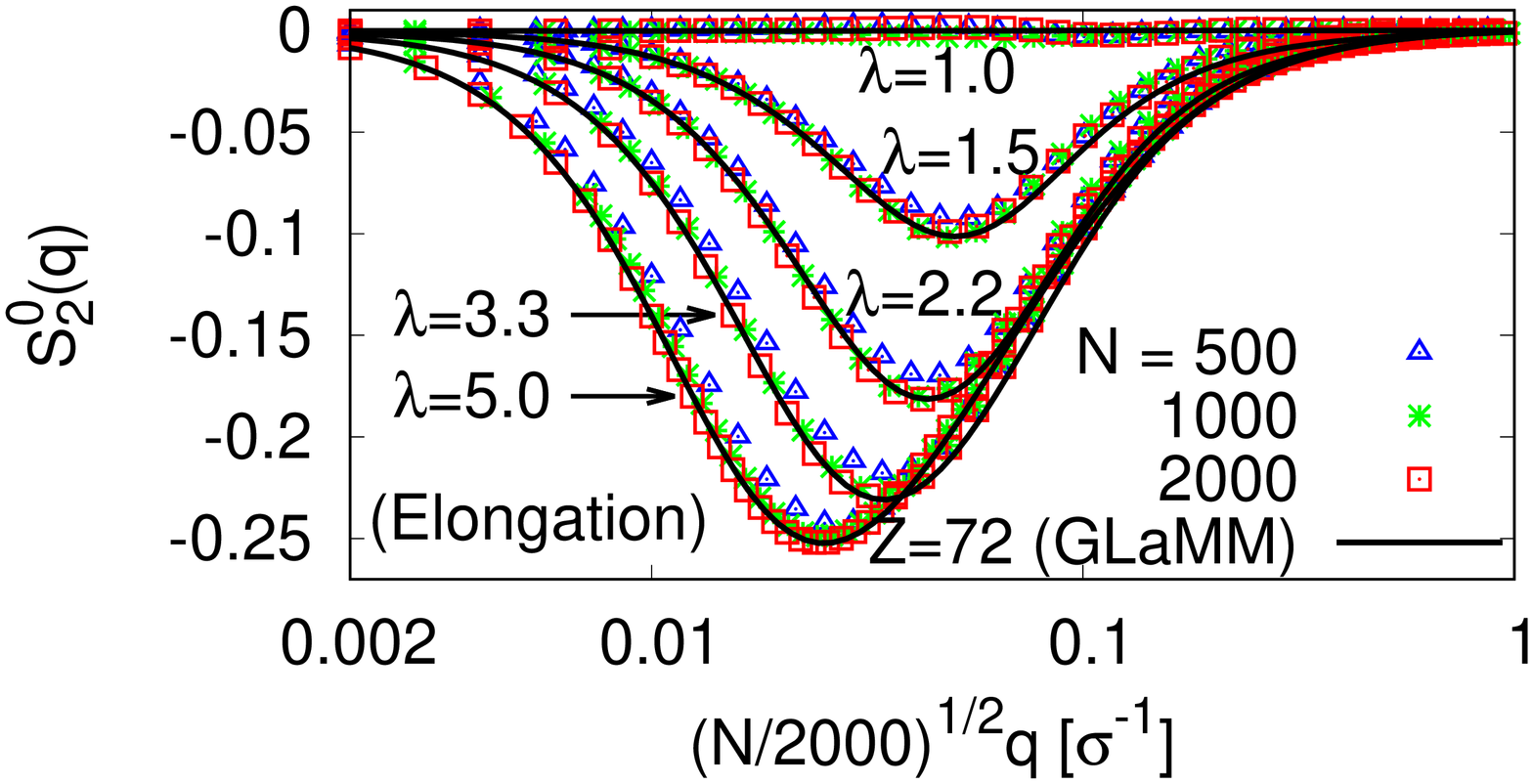} \\
(b)\includegraphics[width=0.25\textwidth,angle=270]{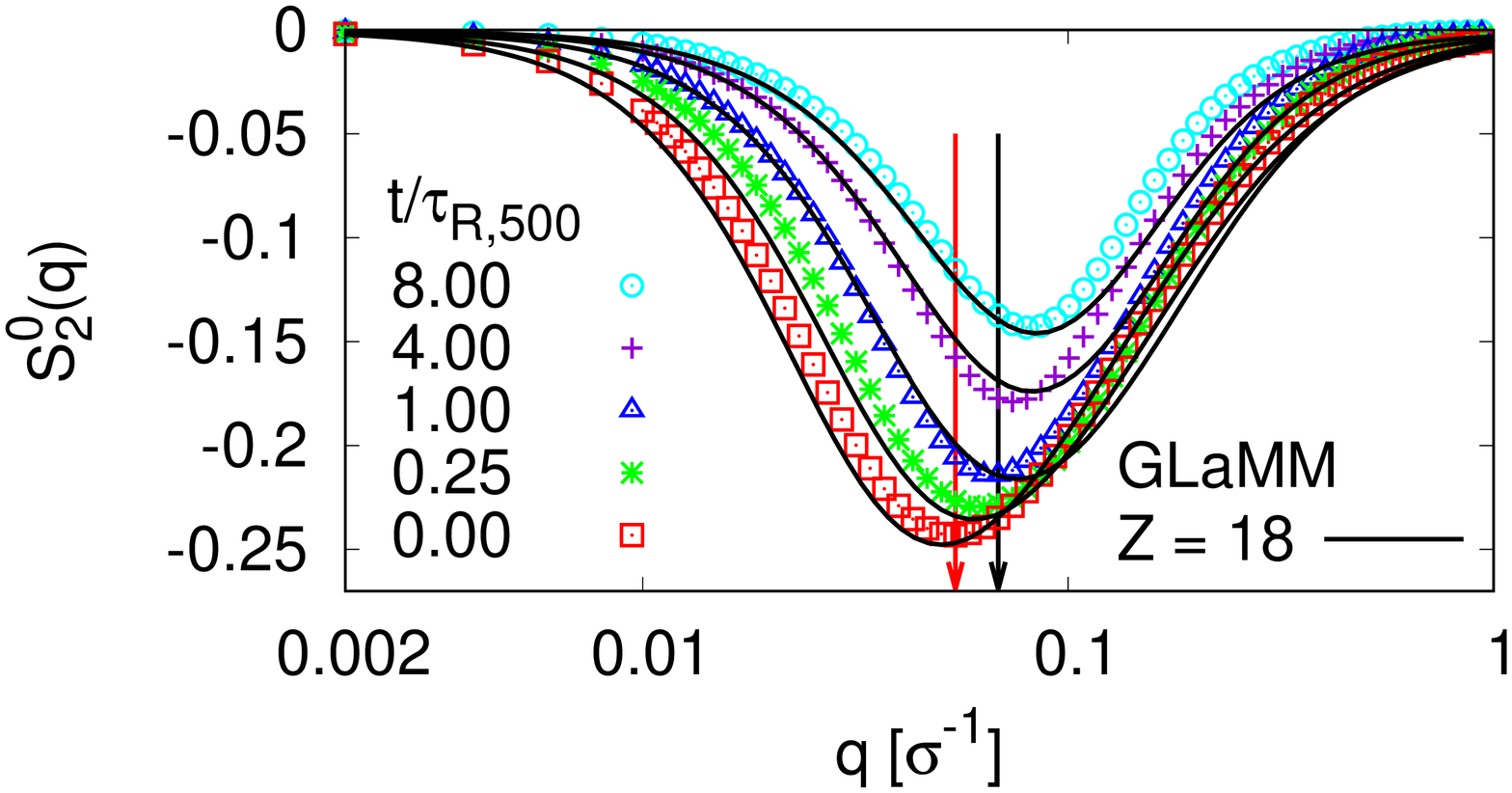} \\
(c)\includegraphics[width=0.25\textwidth,angle=270]{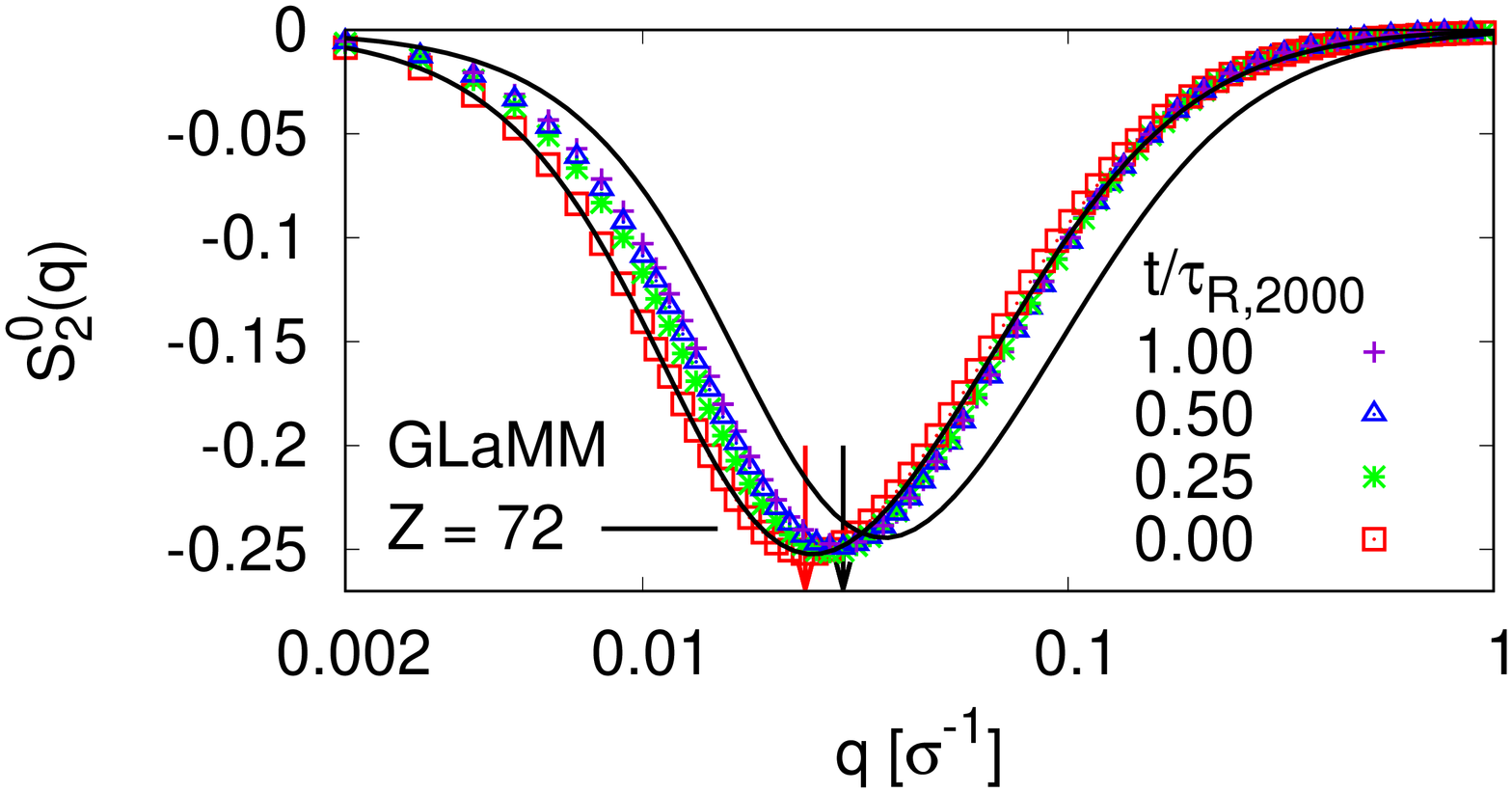}
\caption{Leading anisotropic term of the single chain structure factor,
$S_2^0(q)$, plotted versus $q$ for elongated polymer melts (a)
and upon subsequent relaxation for chain sizes $N=500$ (b), and $2000$ (c).
In (a) five values of stretching ratio $\lambda$, and three
different chain sizes $N$ are chosen, as indicated. 
In (b)(c) data are for several subsequent relaxation times $t/\tau_{R,N}$ after 
stretching, as indicated.
The predicted results from the GLaMM model are shown by 
black curves for all cases in (b), but only
for $t/\tau_{R,2000}=0$ and $1$ in (c).
Note that for the GLaMM model $q$ here is rescaled to $0.62qN_e^{-1/2}$
such that simulation data and theoretical predictions are coincidental in (a).}
\label{fig-rel-sqcqall2}
\end{center}
\end{figure}

Experimentally scattering functions are more easily accessible. The normalized 
single chain structure factor \textcolor{black}{$S({\mathbf q})$}
will easily detect any anisotropy after deformation. 
As for $R_g$ we distinguish $\textcolor{black}{S_{||}}(q_{||}=q_x)$ where the wave vector 
$\textcolor{black}{\bf q}$ is oriented in the $x$ direction parallel to the stretching 
direction, and $\textcolor{black}{S_{ \perp}}[q_\perp=(q_y^2+q_z^2)^{1/2}]$. 
Note that here we discuss the static structure factor for deformed polymers 
in melts at certain selected relaxation times.
In Fig.~\ref{fig-rel-Sqcxyz}, we present the two components 
$\textcolor{black}{S_{||}}(q_{||})$ 
and $\textcolor{black}{S_{\perp}}(q_\perp)$ for $N=2000$ in deformed ($\lambda=5$) melts. 
After a large step elongation, $\textcolor{black}{S_{||}}(q_{||})$ and 
$\textcolor{black}{S_{\perp}}(q_\perp)$ 
strongly deviate from ideality.
In the Guinier regime, $ q < {2 \pi/R_g}$, our data are very well described by the decomposed Debye function, as indicated.
With increasing relaxation time, the range over which the ideal behavior holds 
slowly extends, however, it remains still far from that of ideal chains. 
As expected from $R_g$ the chain retraction as observed for $N=2000$, clearly 
shows up in the Kratky plot of the structure factor $S_{\perp}(q_\perp)$ [Fig.~\ref{fig-rel-Sqcxyz}(b)].
The peak height first increases for times up to $t/\tau_{R,2000}=0.3$, the time
where $\langle R_{g,\perp}^2 \rangle$ reaches a minimum 
(see Fig.~\ref{fig-rel-Rgxyz}).
Then it decreases as $\langle R_\perp^2 \rangle$ turns to increase.
This reduction for $t/\tau_{R,2000}>0.3$, reveals a reduction of the anisotropy. 

To compare this to theoretical predictions we follow previous work of 
Refs.~\cite{Wang2017,Xu2018} and employ 
an expansion with respect to spherical harmonics
to the single chain structure factor \textcolor{black}{$S({\mathbf q})$}. This should reveal the 
relationship between anisotropic chain structure and chain retraction for the 
leading anisotropic term.
To take into account axial symmetry, we choose the polar angle $\theta$ to
be the angle between \textcolor{black}{${\mathbf q}$} and the $x$ axis. 
Then the structure factor is independent of the azimuthal angle $\phi$, 
implying that the expansion with respect to the spherical harmonics 
$Y_\ell^m(\theta,\phi)$
exhibits only terms with $m=0$, $Y_\ell^0(\theta)$. In practice, we simply
set $\phi=0$ and thus obtain $\textcolor{black}{S}(q_x=q\cos \theta, q_z=q \sin \theta)=
\sum_{\ell=0,2,4,\ldots} S_\ell^0(q)Y_\ell^0(\theta)$, where odd $\ell$ values
do not occur for reasons of mirror symmetry. Focusing on the leading order
anisotropy, we present in Fig.~\ref{fig-rel-sqcqall2} the coefficient
$S_2^0(q)$, for polymer melts of our three
different chain sizes within the elongation process at five selected strain 
values $\lambda$, and during the relaxation process at fixed
$\lambda=5.0$. 
Since polymer chains deform affinely, and
$q \propto 1/\langle R_g^2 \rangle^{1/2} \propto 1/N^{1/2}$,
we rescale $q$ to $(N/2000)^{1/2}q$ in Fig.~\ref{fig-rel-sqcqall2}(a).
As expected, we observe a nice data collapse for chains of different $N$.
With increasing $\lambda$, the anisotropy of deformed polymer chains 
in a melt is enhanced. 
The differences between the gyration radii along the $x$ and $z$ axes
become more pronounced, resulting in a horizontal shift of $S^0_2(q)$ to smaller values of $q$. 
Meanwhile, the orientation anisotropy becomes stronger,
i.e. the minimum of $S^0_2(q)$ becomes deeper.
\textcolor{black}{So far the agreement between the GLaMM model and the simulation is excellent}. 

As the deformed chains start to relax, the situation changes.
For all cases [Fig.~\ref{fig-rel-sqcqall2}(b) and \ref{fig-rel-sqcqall2}(c)],
we indeed see a horizontal shift of $S^0_2(q)$ to larger values of $q$
due to the shrinkage of chains within the initial relaxation up to about the Rouse time. The minimum of $S_2^0(q)$ becomes more shallow, depending  
on the number of entanglements $Z$.
For better illustration, the minima at $t/\tau_{R,N} \approx 0$ and $1.0$
are indicated by arrows. As observed directly by analyzing $R_g$ the GLaMM model 
seems to reproduce this relaxation better for small $Z$, indicating significant 
deviations from the GLaMM relaxation mechanisms with increasing chain length.
\textcolor{black}{Results of the higher order terms $S^0_4(q)$ and $S^0_6(q)$ 
are shown in Supplemental Material~\cite{SM}}.

In summary, both results of the radius of gyration and the one-dimensional 
structure factor of deformed melts indicate that chain retraction in all 
directions sets in during initial relaxation before reaching the Rouse time.
We find that the signature becomes more pronounced with an increasing number of 
entanglements $Z$ as predicted by the GLaMM model.
Such an effect was not observed in Refs.~\cite{Wang2017,Xu2018}. Our data 
indicate that there the number of entanglements $Z$ is not big enough and/or the applied strain is not large enough; 
i.e., the stretch ratio $\lambda=1.8$ is too small.
We have also shown that during the relaxation process up to the Rouse time, 
the leading anisotropic term of the single chain structure factor follows a 
similar pattern as predicted by the GLaMM tube model.
Beyond the initial agreement with the GLaMM model at short 
times significant deviations have been observed for larger times. This 
relaxation retardation needs further investigation, as it points to different, 
not yet understood relaxation pathways in the nonlinear viscoelastic regime 
of highly entangled polymer melts.

\section{acknowledgement}
We are grateful to B. D\"unweg for a critical reading of the manuscript.
H.-P. H. thanks R. Graham for helpful discussions on the GLaMM tube model.
This work has been supported by the European Research Council under the European
Union's Seventh Framework Programme (FP7/2007-2013)/ERC Grant Agreement
No.~340906-MOLPROCOMP.
We gratefully acknowledge the computing time granted by the John von
Neumann Institute for Computing (NIC) and provided on the supercomputer JUROPA
at J\"ulich Supercomputing Centre (JSC),
and the Max Planck Computing and Data Facility (MPCDF).

%\bibliography{Ref_2018}
%\end{document}

%
\end{document}